# Voluntary phantom hand and finger movements in transhumeral amputees could be used to naturally control polydigital prostheses

Nathanaël Jarrassé, Caroline Nicol, Florian Richer, Amélie Touillet, Noël Martinet, Jean Paysant and Jozina B. De Graaf

*Abstract*—An arm amputation is extremely invalidating since many of our daily tasks require bi-manual and precise control of hand movements. Perfect hand prostheses should therefore offer a natural, intuitive and cognitively simple control over their numerous biomimetic active degrees of freedom. While efficient polydigital prostheses are commercially available, their control remains complex to master and offers limited possibilities, especially for high amputation levels. In this pilot study, we demonstrate the possibility for upper-arm amputees to intuitively control a polydigital hand prosthesis by using surface myoelectric activities of residual limb muscles (sEMG) associated with phantom limb movements, even if these residual arm muscles on which the phantom activity is measured were not naturally associated with hand movements before amputation. Using pattern recognition methods, three arm amputees were able, without training, to initiate 5-8 movements of a robotic hand (including individual finger movements) by simply mobilizing their phantom limb while the robotic hand was mimicking the action in real time. This innovative control approach could offer to numerous upper-limb amputees an access to recent biomimetic prostheses with multiple controllable joints, without requiring surgery or complex training; and might deeply change the way the phantom limb is apprehended by both patients and clinicians.

## I. INTRODUCTION

A perfect hand prosthesis should offer amputees a natural, intuitive and cognitively simple control over numerous biomimetic active degrees of freedom (DoF) [1, 2]. But there is an evident gap between devices and control approaches. Efficient polydigital prostheses such as the iLimb hands by Touch Bionics [3] or the Michelangelo hand by Ottobock [4] (see [5] for a comparative review) are already commercialized, but are only available with limited control approaches such as a switchable catalog of postures chosen by co-contraction impulses. Researchers have been working for numerous years in developing alternative approaches to enhance controllability over active prosthetics and to offer direct control of each active Dof. Approaches like Targeted Muscle Reinnervation (TMR) [6] or brain computer interfaces (BCI) [7] are promising solutions, but they require heavy surgery with still reduced performances, especially for BCI.

Non-invasive approaches, based on the use of electrode arrays placed over the residual group of muscles and pattern recognition algorithms, have made huge progress within the last decades [8]. This last approach has been recently reported to offer forearm amputees a control over a polydigital hand prosthetics via decoding of finger muscle activities with an array of electrodes placed over the residual forearm [9].

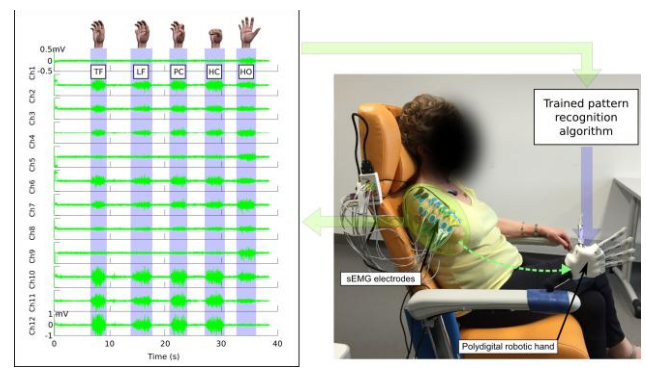

Figure 1. Global view of the setup along with the myoelectric activity associated to the voluntary control of phantom hand (measured with 12 sEMG electrodes placed over the residual limb of one participant). Phantom hand movements are named the following way: TF stands for Thumb Flexion, LF for Little finger Flexion, PC for Pinch Closing, HC for Hand Closing and HO for Hand Opening.

Although not often explicitly mentioned, such control over remaining muscles in the residual forearm might be related to the "mobile phantom limb" phenomenon. Most amputees report having a phantom limb, i.e., the perception that their lost limb is still present and attached to their body [10], and many of them describe the capacity of voluntary mobilization of this phantom limb (85% of arm amputees, according to a recent epidemiological study we performed on mobile phantom limbs). Since phantom movements cannot be observed, they are often confounded with motor imagery of the lost limb. Yet, recent studies show that movement control of the phantom limb is different from motor imagery [11,12], an important argument being that, contrary to motor imagery, muscle activity is always present in the residual limb during phantom movement execution, with muscle activation patterns that are not random but seem to depend on the type of phantom movement [13,14]. This is not so much surprising for forearm amputees for whom residual muscles were already involved in finger and wrist movements before the amputation. Yet, it was also found for transhumeral amputees for whom the residual limb muscles had never been involved in finger and wrist movements before amputation. Therefore, decoding voluntary phantom hand movements from residual upper limbs differs from that from residual forearms.

N. Jarrassé and F. Richer are with Institute of Intelligent Systems and Robotics, ISIR CNRS-UMR 7222, INSERM U1150 Agathe-ISIR, Sorbonne University, UPMC Univ Paris 06, Paris, France (e-mail: jarrasse@isir.upmc.fr).

C. Nicol and J. De Graaf are with the Institute of Movement Sciences, UMR 7287 – CNRS & Aix-Marseille University, Marseille, France (e-mails: caroline.nicol / jozina.de-graaf@univ-amu.fr).

A. Touillet, N. Martinet and J. Paysant are with the Louis Pierquin Centre of the Regional Institute of Rehabilitation, Nancy, France (e-mail: amelie.touillet / noel.martinet / jean.paysant@ugecamne.fr).

In the present study, we evaluated the possibility to decode phantom fingers activity, to offer upper arm amputees a dexterous control of polydigital prosthetics by using the sEMG activities recorded over the arm residual limb while mobilizing the phantom limb. Our approach is based on classical sEMG pattern recognition methods used in the present context to classify phantom finger and hand movements in order to have these phantom movements in real-time mimicked by a polydigital hand prosthesis. We were recently able to classify online phantom movements of elbow, wrist and hand based on the associated sEMG recordings from the residual upper-arm muscles, with an average successful classification rate over 80%, and to offer participants a real-time control over a graphical user interface through their phantom limb mobilization [15]. These preliminary results pushed us to believe that such approach could be transferred to the decoding of phantom finger activity, and used to a novel control approach of polydigital prostheses for the misconsidered upper arm amputees.

## II. METHODS

### A. Participants

Three volunteers participated in the study (2 men, 1 woman, respectively 33, 62 and 77 years old), recruited at the Institute of Physical Medicine and Rehabilitation. All patients were healthy except that they had one unilateral transhumeral amputation following a traumatic accident. The brachial plexus of one participant (P2) was affected by the accident preceding the amputation. Before the accident all patients were right-handed. The inclusion criteria were a good control ability of the mobile phantom limb as well as repeatable phantom limb movement of a minimal 20° amplitude, an absence of phantom limb pain, and the availability of the participant during the recording period. The study was approved by the Ethical commission of the Institute of Physical Medicine and Rehabilitation, and performed in accordance with the Declaration of Helsinki. The participants provided written informed consent to take part in the study.

The actual voluntary mobilization of the phantom limb was precisely defined using a questionnaire that intended to make clear distinctions between sensations in the residual limb, phantom pain, phantom sensations as well as between mobility of the residual limb and of the phantom limb. Patients' demographic and clinical data are reported in the Table 3. The participant (P2) with an affected brachial plexus pre-amputation suffered from a prosthetic fitting failure. The absence and reduced control over his biceps and triceps, respectively, required the prothesist to place the electrodes of his classical myoelectric prosthesis on the trapezoidal muscles. All three participants were able to perform voluntarily movements of thumb, index and major fingers along with whole hand opening and closing actions. Table 4 presents the list of the hand movements that each subject was able to perform.

### B. Experimental setup

We used the BCI2000 software suite to develop the global control architecture on a desktop computer running Windows 7 (Intel Core i5-4690K (3.5 GHz) with 16 Go DDR3). BCI2000 is a general-purpose software suite designed for brain-computer interface (BCI) and was used here to run in parallel three principal modules: one acquisition driver to acquire the sEMG data (at a frequency of 1kHz), one Matlab© classification algorithm script (executed every 128ms), and one driver (C++) to control the Prensilia IH2 Azzura robotic hand through a serial port.

An electrophysiological recording system (Eegosports, ANT-Neuro©, the Netherland) with 24 bipolar channels and 24-bit resolution was used to record sEMG activity from the participant's residual limb muscles at a sample rate of 1 kHz. Because of the variety in residual limb length and muscle anatomy, the scheme of electrode placement had to be adapted for each participant. Twelve pairs of active sEMG electrodes (Ag/AgCl snap bipolar electrodes with a 1.25cm-diameter circular contact) were used for each participant to record activity on various parts of the residual muscles where slight contraction could be felt by hand during phantom movements. Placement for each participant is shown in Figure 2. Twelve couples of electrodes were placed over the residual biceps, triceps, deltoid and sometimes trapezoidal and grand pectoralis muscles. No specific skin preparation was used before placing the active electrodes over the residual limb. The recorded sEMG signals were then filtered with a [10 Hz ; 400 Hz] third-order bandpass Butterworth filter.

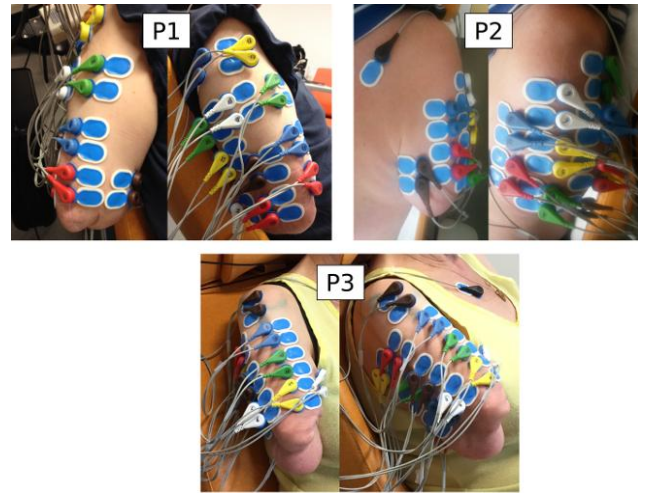

Figure 2. 12 electrode pairs were placed on the residual limb of each of the three participants. Two views for each participant.

In this study, we used a linear discriminant analysis (LDA) classifier [16] running on the Matlab© Engine within the BCI Toolbox suite. An LDA classifier was chosen because it is based on a simple and robust statistical and thus computationally efficient approach. Features were computed from the sEMG using a 512-ms sliding analysis window with a 128-ms overlap between successive windows. The first 4 autoregressive coefficients (AR), the root mean square (RMS) value and the sample entropy of the sEMG were extracted for each channel and used to create the feature vector. The confidence value of the classifier was used to filter the algorithm output: if classification confidence was below 90%, no order was sent to the robotic hand.

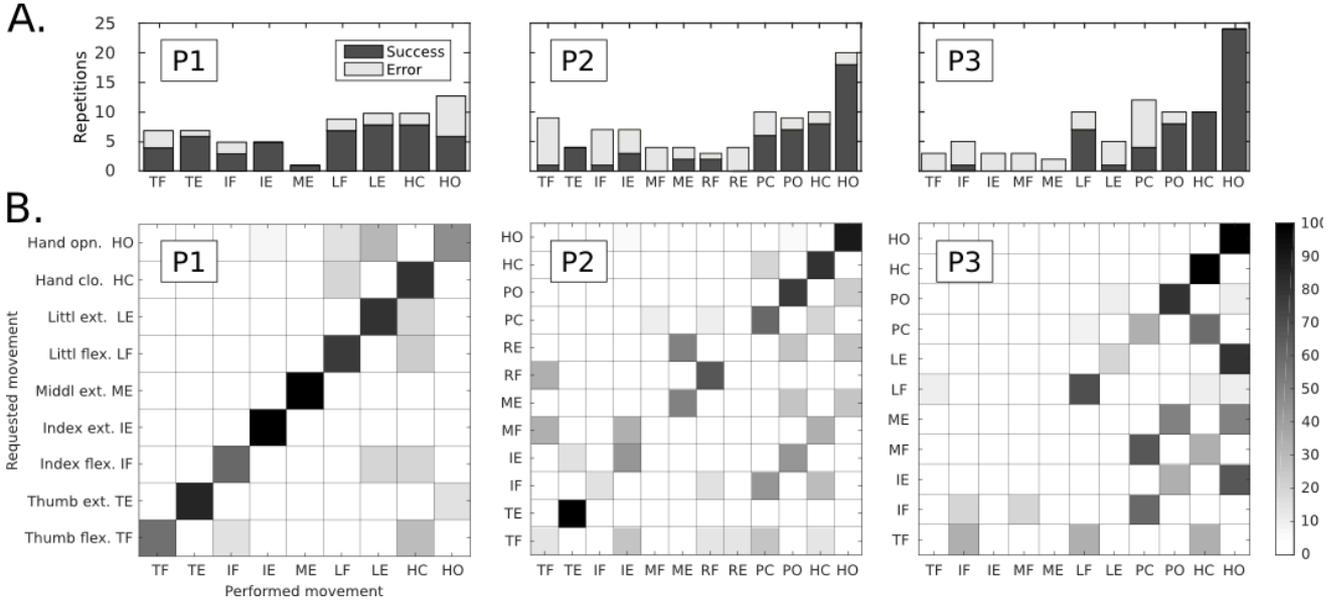

Figure 3. Control of all possible phantom movements (Test 1).A: Number of repetitions performed during Test 1 for each possible type of phantom movement and the associated ratio of successful classification (in black). The results are shown for each participant individually (P1-P3). B: Confusions matrices for each of the three participants of our multi-feature set for Test 1.The confusion matrix color scale increases from white to black as a function of increasing classification rate. Remark: participant P1 was not able to perform MF, RF, RE, PC, PO; P2: LF and LE; P3 could not perform TE.

We use a right 5-DoF Prensilia IH2 Azzurra robotic hand. Position control of the fingers is used to generate a velocity-like control: each time a movement class is detected with sufficient confidence, an angular position displacement (5°) is sent to the concerned finger or group of fingers. A specific filter is added to avoid punctual changes in the classification output to be sent to the hand (a given movement has to be by the robotic hand).

*C. Protocol*

The experiments are organized in two testing series: Test 1 with a control of all possible phantom movements and Test 2 with only preferred movements. Each test consisted of two steps: first training of the classification algorithm, and then testing of its performances with online control of the polydigital robotic hand by the participant. The overall duration of the experimental session was approximately 2 hours, including the placement of electrodes. Both tests were videotaped.

The patient was comfortably seated in a specific dedicated chair, fitted with armrests and a head rest. The robotic hand was attached to one of the arm-rests of the chair, and placed in a chosen position to be as much as possible symmetric with the intact arm lying on the other armrest. The first test started by verifying the previous description of their phantom hand movements and the effort required for their execution in order to determine the phantom movement testing order. Yet, when during the training phase of Test 1 the participant mentioned that he/she was able to perform an extra gesture (which happened sometimes), the latter was added to the catalog of possible movements, even if the participant was not fully sure of its repetition capability.

Once the electrodes were placed, Test 1 started by a training phase during which the sEMG activities of the residual limb were recorded. The participants were asked to perform 2 repetitions of each possible phantom movement with a few minutes of rest in-between. The experimenter was in charge of verbally asking the subject to perform tasks and therefore was in charge of the rhythm of the performance. No instruction was given about the velocity of the gesture or its amplitude. Then, the recordings were manually tagged offline with the movement type (from Thumb Flexion TF to Hand Opening HO) by the experimenter through a dedicated interface. This did not last more than 5 minutes. Once these recordings were obtained, the data were analyzed and assembled to constitute the training set of data to be used in the testing phase (online classification of the phantom limb activity and real-time mimicking by the robotic hand). Similar training was performed at the beginning of Test 2 with only some preferred movements. On the average, the inter-test time delay was 10 to 15 minutes.

During the testing phase, the participants were asked by the experimenter to perform a list of randomized movements of their phantom limb (all interspersed with pauses) while the classification algorithm was trying to identify these movements (thanks to the set of training parameters previously obtained) and to reproduce them in real-time with the robotic hand.

### III. RESULTS

We performed experiments on three upper-arm amputees with voluntary motor control over their phantom limb. The specificity of our protocol was that the participants were neither trained nor used to mobilize their phantom limb (apart from P1 who reported regularly mobilizing his phantom little finger to release unpleasant contractions in the residual limb). This resulted sometimes in surprise, fatigue and tensions, and even revealed some for-the-participants-unknown mobilization capabilities (see Protocol in Methods). Therefore, the number and types of possible phantom movements varied between the participants (see [17] for a characterization of phantom movement kinematics).

Twelve electrode pairs were placed over the residual limb (i.e., the upper arm). First, a very short training phase of the pattern recognition algorithm was done based on only two repetitions of each possible phantom movement (see Protocol in Methods).Then, for the subsequent control tests, a polydigital robotic hand was attached to the armrest of their chair at the approximate location of a normal prosthetic hand (see Fig. 1). The two tests consisted of the experimenter asking them to perform a pseudo-random series of given phantom movements among the catalog of their possible movements.

In the first test (see Methods) the participants were asked to perform a pseudo-random sequence of phantom movements among their individually defined catalog of possible movements. Fig.3A shows, for each participant, the variation in the number of repetitions of each type of phantom movement as well as the associated ratio of successfully mimed movements by the robotic hand. When considering the whole range of phantom movements (i.e., 9-12 different movements), the obtained individual rates varied from 57.1 to 71.6% of successful reproductions. Yet, some movements being never correctly recognized by the classifier, practically, the participants were able to control 5 to 8 different gestures with this control approach.

It can be seen that the catalog of possible movements as well as the number of repetitions for a given movement varied between the participants. Also, the rate of successfully mimed movements by the robotic hand was rather variable over the different types of movements (Fig. 3A). The size of the catalog of possible phantom limb movements is probably influencing the recognition by the classifier of the phantom limb actions. The bigger the catalog is, the more frequent the confusion between gestures can be for the pattern recognition algorithm (see Fig.3 B). The interesting result is that even if the distinction between finger actions is not perfectly clear, the distinction between flexion/closing and extension/opening is generally obtained.

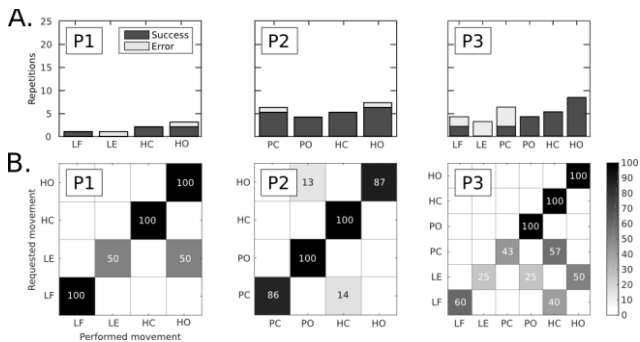

Figure 4. Control of preferred phantom movements (Test 2).A: Number of repetitions performed for each movement among the catalog of preferred movements, and associated ratio of classification success. B: Confusion matrices for each of the 3 participants for our multi-feature set for Test 2 (preferred /easiest movements). The confusion matrix color scale increases from white to black as a function of increasing classification rate.

Since all phantom movements were not performed with the same facility, we conducted a second test with a reduced set of preferred/easiest movements (i.e., 4 to 6 gestures for each participant). The inter-individual variations between the repetition numbers of each movement are shown in Fig.4A along with the associated distribution of successfully mimed movements by the robotic hand. Confusion matrices between the phantom limb gestures performed by each subject (asked by the experimenter) and the movement mimed by the robotic hand are shown in Figure 4.B. This reduction in the size of the movement catalog to only the preferred movements led to a clear increase of the successful reproduction rates for two participants (P2 increasing from 57.1 to 92.3 % and P3 from 63.2 to 75 %, see Table 2).

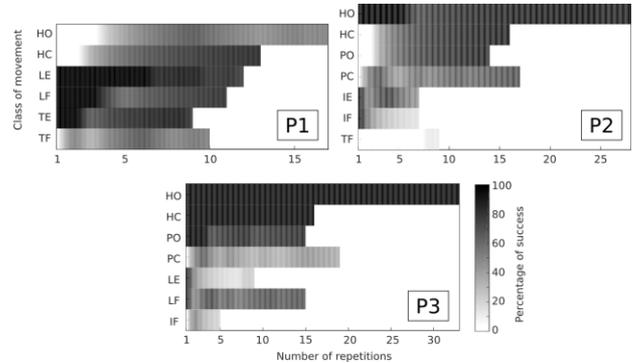

Figure 5. Classification performance as a function of the number of repetitions. This figure was created by only considering the movements that were performed a total of at least 5 times in the two tests. The color scale increases from white to black as a function of increasing classification rate.

The global results suggest that two phenomena may interfere during the test: practice and fatigue effects. For all three participants, when the control of a phantom limb movement was easy for them, the sEMG activation pattern had a high probability to remain stable along time, and thus to be successfully mimed for several repetitions by the robotic hand. Indeed, the control of some movements was found to be improved across trials or at least stable among task repetitions (see Fig. 5). On the other hand, the activation patterns of phantom finger movements that hadn't been executed regularly and/or for a long time were more effortful and probably changed during the intensive testing series such that the classifier could not recognize the movement anymore.

## IV. Discussion

The first hypothesis, concerning the high control dimensionality allowed by this control method, is well supported by the present results. The control dimensionality of this phantom limb associated muscle activity is high since subjects were able to initiate 4 to 8 different movements, with a successful classification rate of 78%. Such approach is definitively promising for offering arm amputees a full and functional control over commercially available polydigital hand prosthetics, similarly to what is being studied in research laboratories for forearm amputees. Classification performances based on sEMG activity of forearm amputees are generally slightly higher (80%) [9], but the sEMG signals are recorded by arrays of electrodes placed over the partly preserved forearm muscle groups whereas in the present study we recorded over upper arm muscles. Moreover, the literature reports a reduced catalogue of 4 to 6 phantom wrist and hand (opening/closing) movements [18] and rarely of individual finger movements as we do in the present study. For upper-arm amputees who want to increase the number of active prosthetic DoF, our approach actually constitutes the only alternative to approaches requiring surgery like TMR. Indeed, similar to TMR, the mapping between the input signals and the output controlled joints is direct and intuitive: the phantom joint movement is simply mapped into similar joints on the robotic hand.

Our second hypothesis was that the present prosthetic control approach (real-time mimicking of the phantom hand by the robot) is intuitive and doesn't need prior learning. This hypothesis is confirmed by our preliminary results showing the possibility to control, without any training or surgery, dexterous prosthetics based on sEMG muscle activities associated to voluntary phantom limb movements. It is fundamental to underline that the sEMG activities were measured over the residual limb, and thus on muscles that were not involved in wrist, hand and finger movements before the amputation. Furthermore, even if this approach relies on the measurement of sEMG signals, it does not require a fine control of the EMG activation levels but, instead, the control of a movement, which is much more natural for humans. Indeed, human motor control is dedicated to movement and force control rather than to selective control of individual muscle activity [19]. This is why even simple and generic myoelectric control approaches for hand prosthetic control based on bicep/triceps sEMG activations require long training in order for the patient to be able to generate single-muscle activation and avoid "automatic" co-activations.

Although our control approach clearly does not need any prior learning, practicing phantom movements seems to stabilize the associated sEMG patterns, which eases the recognition by the classifier. This is indeed supported by the improvements for 2 participants when they performed their preferred movements that generally are the ones that they practice to relax their residual limb muscles or just "to play with the phantom". Thus, while patients do not have to learn a new mapping between muscle contractions and robotic degrees of freedom, regular practicing phantom movements might still improve their control and the associated classification performances. Therefore, patients and the medical community should go beyond the usual skepticism concerning phantom movements and practice them regularly.

The third hypothesis was that, as long as the patient is able to mobilize the phantom limb, the condition of the residual limb as well as the age of participants are no limiting factors. Elderly persons often show a decline in physical and cognitive capacities, limiting their learning possibilities. Yet, encouraging results were obtained with the oldest participant (P3, 77 years, amputated for 14 years), which underline the simplicity of this intuitive control approach. Moreover, patients with affected plexus bracchi (P2 in our study) generally have trouble to be fitted with conventional myoelectric prosthetics because of residual limb muscle paralysis and/or altered and weak sEMG signals. Yet, their partly-paralyzed residual limb still exhibits myoelectric activity patterns encoding the phantom limb movements. For instance, while P2 had been trained to control his conventional myoelectric hand prosthetics by activating his trapezoid muscles (due to the partial paralysis of the biceps and triceps muscles), he was able to intuitively control the polydigital hand by the sEMG activity recorded from his biceps and triceps muscles, even if he couldn't contract them voluntarily other than by mobilizing his phantom limb.

The major limiting factor on the control dimensionality remains the controllability of the phantom limb: if only a very limited number of movements can be performed, the interest of our control approach is reduced. Nevertheless, executing a phantom movement remains more natural than learning to control individual muscle activation levels. Moreover, training of the phantom movements could, in addition to reducing phantom limb pain (PLP) [20], help unlocking extra phantom limb movements, as we observed in the present study. So even with a reduced set of possible phantom limb movements, since patients tend generally to perform at least phantom hand opening and closing, this approach could even be a good alternative to control simple conventional 1 or 2 DoF myoelectric hands.

In conclusion, the present results are very encouraging, especially since current solutions for transhumeral amputees who want to control a polydigital hand is to undergo either TMR surgery or an intensive training in order to learn an indirect mapping between muscle contractions and different prosthetic joint actions, and/or to explore a catalog of postures with co-contraction signals. The present intuitive control approach could deeply impact the use and appropriation of upper-limb prosthetics and might offer transhumeral amputees access to recent technological advances by:

• offering new perspectives to amputees with already varied and selective voluntary control of their phantom limb: they can, without either surgery or heavy training and cognitive load, intuitively obtain a more selective control leading to an increased number of active prosthetic DoF, even of individual fingers;

• simplifying the control for other patients: upper-arm amputees with reduced phantom limb mobility and fitted with simple conventional myoelectric hand prosthesis (opening/closing of the hand and active prono/supination of the wrist) could benefit from this control mode and obtain a more intuitive and non-sequential control.

• offering amputees with a much deteriorated residual limb anatomy and function (affected brachial plexus) a supplementary chance to be fitted with myoelectric prostheses.

Finally, these results could deeply change the way the phantom limb condition is apprehended by both patients and clinicians. Indeed, the phantom limb still makes both patients and clinicians feel uncomfortable [21], leading to completely ignored mobile phantom limbs [22]. So far, the mobility of the phantom limb has no use for rehabilitation of upper-arm amputee and is often considered by the amputee and/or the family as a sign of non-acceptance of the limb loss. Also, interference between phantom limb movements and conventional prosthesis control is often evoked. Yet, since the prostheses will be controlled by the phantom limb, this will clearly not be a problem anymore. And the important positive co-effect of the present approach will be the acceptance, and even the encouragement to consider as useful for daily life, the mobility of the phantom limb by patients, their family and rehabilitators.

## V. Conclusion

A conclusion section is not required. Although a conclusion may review the main points of the paper, do not replicate the abstract as the conclusion. A conclusion might elaborate on the importance of the work or suggest applications and extensions.

### Acknowledgment

We warmly thank the participants who so willingly gave us several hours of their time. We are also very grateful to the medical staff and the Prosthetics Department of the IRR Nancy.

This project was supported by the CNRS (project Subilmaof Défi AUTON 2016; PEPS INS2I JCJC MOFACO 2015), the ANR (project PhantoMovControl ANR-15-CE19-0008-02), the Labex SMART (ANR-11-LABX-65) and by the Region Provence-Alpes-Cote d'Azur (Project ExplorAmp 2012, no 2012-07072).